\documentclass[%
 reprint,
superscriptaddress,
 amsmath,amssymb,
]{revtex4-2}

\usepackage{natbib,hyperref}

\usepackage{graphicx}
\usepackage{dcolumn}
\usepackage{bm}
\usepackage{hyperref}

\usepackage{subcaption}

\newcommand{\ra}{\rightarrow}

\begin{document}

\title{Revisiting relativistic electrically charged polytropic spheres}

\author{Andr\'es Ace\~na}
\affiliation{Instituto Interdisciplinario de Ciencias B\'asicas, CONICET, Facultad de Ciencias Exactas y Naturales, Universidad Nacional de Cuyo,
Mendoza, Argentina}

\author{Bruno Cardin Guntsche}
\affiliation{CONICET, Instituto Balseiro, Comisión Nacional de Energía Atómica, Bariloche, Argentina.}

\author{Iv\'an Gentile de Austria}
\affiliation{Facultad de Ciencias Exactas y Naturales, Universidad Nacional de Cuyo, Mendoza, Argentina.}

\begin{abstract}
We revisit the problem of the structure and physical properties of electrically charged static spherically symmetric solutions of the Einstein-Maxwell system of equations where the matter model is a polytropic gas. We consider a relativistic polytrope equation of state and take the electric charge density to be proportional to the rest mass density. We construct the families of solutions corresponding to various sets of parameters and analyze their stability and compliance with the causality requirement, with special emphasis on the possibility of constructing black hole mimickers. Concretely, we want to test how much electric charge a given object can hold and how compact it can be. We conclude that there is a microscopic bound on the charge density to rest mass density ratio coincident with the macroscopic bound regarding the extremal Reissner-Nordst\"om black hole. The macroscopic charge to mass ratio for the object can exceed the corresponding microscopic ratio if the object is non-extremal. Crucially, the only way to obtain a black hole mimicker is by taking a subtle limit in which an electrically counterpoised dust solution is obtained.
\end{abstract}

\maketitle

\section{Introduction}

In the present article we consider static, spherically symmetric configurations of charged polytropic matter in the context of General Relativity. The study of non-vacuum solutions of the Einstein-Maxwell system of equations has a long and fruitful history, in particular because they provide interior matter solutions for the Reissner-Nordstr\"om (RN) metric. As the interior solutions need to be smoothly joined to the electro-vacuum exterior, the total mass, $M$, and the total charge, $Q$, satisfy the same requirement as the RN black hole for not representing naked singularities, $Q\leq M$, being the extremal case $Q=M$. Also, to be a regular solution and not a black hole, the radius of the object, $R$, has to be larger than the corresponding black hole radius, $R_+ = M+\sqrt{M^2-Q^2}$.

More than 50 years ago the generalization for charged matter of the Tolman-Oppenheimer-Volkoff (TOV) equations of hydrostatic equilibrium was first presented \cite{Bekenstein:1971ej}. There, also the time evolution equations where derived and it was argued that pair creation and high conductivity at the core would discharge a neutron star, and that therefore electric charge inclusion was not necessary to discuss such compact objects. Nonetheless, the question was raised as to whether electric charge may prevent the total collapse of a charged ball. We consider that this question has not yet been completely resolved. Being the TOV equations complicated enough, a natural step to analyze possible solutions is to integrate them numerically, which was done in \cite{Zhang1982}. As matter model it was considered  a completely degenerate Fermi gas and the charge density was set proportional to the energy density. The main conclusion was that there is a maximum charge fraction above which there are no stable configurations, and that said upper bound is far below the macroscopic charge of extremal black holes. Below this bound the families of solutions behave similarly to solutions without charge. In \cite{Anninos:2001yb} also the TOV equations were numerically integrated and a stability analysis was performed. A polytropic equation of state was used and the charge distribution was prescribed \textit{a priori}. The main interest was to determine if relativistic charged spheres can form extremal black holes. The conclusion attained was that they can not, as the spheres became unstable before the extremal limit is reached. Even so, there were hints that if a very stiff equation of state is considered, corresponding to an extremely high polytropic exponent, then the extremal limit can be approached. Again considering a polytropic equation of state with respect to energy density but assuming that the charge density is proportional to energy density, in \cite{Ray:2003gt} the charged solutions with polytropic exponent $\frac{5}{3}$ were constructed. Although it was argued that a relevant amount of charge is not possible for realistic objects, a mechanism is proposed for the formation of charged black holes. With emphasis in the so called quasiblack hole limit, in \cite{Arbanil:2013pua} the TOV equations were integrated numerically and families of solutions were constructed. The polytropic relation was taken with respect to energy density and the charge density was set proportional to energy density. Although the speed of sound was taken into account, it was not considered as determinant when taking the quasiblack hole limit. Later, in \cite{Arbanil:2017huq}, the same type of analysis was done using the relativistic polytropic equation of state, where the polytropic relation was with respect to rest mass density, although again the charge density was proportional to energy density. The results were qualitatively the same, although quantitative differences were obtained for the high density regime. For a description of the concept of quasiblack hole and associated properties the reader is refered to \cite{Lemos:2020ooh}.

Also from an analytic perspective important advances have been done. A substantial effort at classifying spherically symmetric charged solutions was made in \cite{Ivanov:2002jy}, where the classification was done regarding which free functions were prescribed. Many explicit solutions were presented and also known solutions recovered and discussed with regard to its physical properties. Generalizing Buchdahl's proof to the charged setting, in \cite{Giuliani:2007zza} an absolute bound on how compact relativistic charged spheres can be was proved. The main assumptions were that the mass density is a decreasing function of radius while the charge density is increasing. The limit solution is obtained for a constant density sphere, which means that as a fluid it is incompressible. Interestingly, if $Q=M$ the limit is no longer restrictive, as it coincides with the horizon, $R=R_+$. Improving on this result a sharp inequality with less assumptions was proved by Andr\'easson in \cite{Andreasson:2008xw}, and it was shown that infinitely thin shell solutions saturate the inequality.  Constant density charged spheres were investigated numerically in \cite{Arbanil:2014usa}. The charge density was taken proportional to energy density, and the focus was on the Buchdahl-Andr\'easson bound and the quasiblack hole limit. The same problem but from an analytic perspective and restricted to the small charge limit was tackled in \cite{Lemos:2014lza}.

From a fundamental particle physics approach, the spherically symmetric static configurations of neutron stars were studied in \cite{Belvedere:2012uc}, considering the presence of neutrons, protons and electrons. Although only global neutrality was considered, the local charge distribution was obtained from chemical potential equilibrium. Important for us is that with these physically motivated equations of state the resulting objects are far from being black hole mimickers and also charge distribution does not approach extremality. In a similar vein, but without imposing global charge neutrality, the relativistic equilibrium of electrons, protons and neutrons through chemical potential was treated in \cite{Amorim:2021lvr}, with the intention of modelling non-neutral white dwarf stars. There, a bound on total charge was obtained, which is orders of magnitude below extremality.

A related and theoretically relevant development has been the study of electrically counterpoised dust (ECD) spacetimes. Such matter corresponds to a charged perfect fluid without pressure, where the charge and mass densities are perfectly balanced. As the underlying particles have the same mass as charge, then any static distribution is possible, being the gravitational and electrostatic forces always balanced. This was first shown for a system of discrete particles by Majumdar \cite{Majumdar1947} and Papapetrou \cite{Papapetrou1947}, following the work of Weyl \cite{Weyl1917} on axisymmetric spacetimes. If the matter content is restricted to said particles, then to each particle there is an event horizon, which is interpreted as an extremal Reissner-Nordstr\"om (ERN) black hole \cite{Hartle1972}. If instead of black holes one wants to consider regular objects, then the exterior  solution can be matched with static interiors made of ECD \cite{Das1962}, \cite{Varela2003}. The reach of the results presented in \cite{Weyl1917}, \cite{Majumdar1947}, \cite{Papapetrou1947} and \cite{Das1962}, together with the minimum set of assumptions needed to obtain them, was analyzed by De and Raychaudhuri \cite{DeRaychaudhuri1968}. The assumptions made in \cite{Weyl1917}, \cite{Majumdar1947}, \cite{Papapetrou1947}, \cite{Das1962} and \cite{DeRaychaudhuri1968} have been relaxed in several ways, and the results extended to charged perfect fluids with pressure, for example in \cite{Hernandez1967}, \cite{Guilfoyle1999}, \cite{LemosZanchin2017}, or to higher dimensions \cite{LemosZanchin2008}, \cite{LemosZanchin2009}. The fact that any static charge distribution gives rise to a solution of the Einstein-Maxwell field equations has been exploited to test features of General Relativity, such as the relation between charge and mass in the RN solution and the construction of a point charge model \cite{Bonnor1960}, the construction of static objects with unbounded density \cite{Bonnor1972}, to show that unbounded redshifts can be obtained from regular objects \cite{Bonnor1975}, and to discuss the hoop conjecture \cite{Bonnor1998}. In general, the engineered solutions can be made to be as close to the ERN black hole as desired, and this has been analyzed in relation to the bifurcation of solutions \cite{Horvat2005} and it has been shown that such black hole limit is a general feature of ECD solutions \cite{Meinel2011}. This means that a regular ECD object could mimic an ERN black hole as well as desired.

Our main concern here is the possibility of constructing black hole mimickers, and in particular quasiblack holes, made of charged perfect fluid. From the perspective of the known fundamental particles, this endeavor seems hopeless, as there are numerous reasons why the total charge in a compact object is expected to be negligible \cite{Belvedere:2012uc}, \cite{Amorim:2021lvr}. The only option that would give a chance to a macroscopic charge comparable to the gravitational mass of the object is if there would exist a particle with a charge to mass ratio comparable to ECD. Therefore, we take here the charge density of the fluid to be proportional to the rest mass density, interpreting this ratio as the charge to mass ratio of each particle composing the fluid, since in this way the charge inside a fluid element is proportional to the number of charged particles it contains. Also, taking into account that we are in a highly relativistic scenario, we take as equation of state the relativistic polytropic equation of state, where the polytropic relation is with respect to rest mass density. In order to consider the obtained objects to be reasonable physical objects we consider a stability condition and a causality condition. For stability we take the usual criteria that the object, which is part of a family of objects, has to be in the stable branch, where the limit between stable and unstable solutions is given by the maximum in mass with respect to central pressure. The causality condition is that the speed of sound in the fluid has to be below the speed of light. To the best of our knowledge this is the first time that these constitutive equations and conditions have been imposed together. We think this is relevant because the quasiblack hole limit that we analyze is a delicate limit, and we do not want to include unphysical solutions nor exclude relevant ones. Our main results are as follows. Extremally charged solutions can be formed in the sense of $\frac{Q}{M}\approx 1$, but they have large $R$ and $M$ and low pressure, making them in fact faint extended objects. Black hole mimickers and quasiblack holes, in the sense of $R\approx R_+$, can not be constructed using polytropic charged matter. The limit that gives a quasiblack hole is subtle, and entails taking at the same time the charge density to mass density ratio to one, the polytropic exponent to infinity and the central pressure to zero, which in fact means that it is an ECD solution. This shows that non-extremal quasiblack holes can not be formed even in this limit. Also, together with the results in \cite{Acena2021} and \cite{Acena2023}, this makes extremely difficult to consider a physically feasible scenario where the collapse of regular matter leads to the formation of an ERN black hole.

The article is structured as follows. In Section \ref{SystemOfEquations} we present the TOV system of equations and the equations of state that we consider, together with the physical requirements that we impose on the solutions. In Section \ref{FamiliesOfSolutions} we present the families of solutions constructed numerically. In Section \ref{MimickerLimit} we show that the black hole mimicker limit corresponds to ECD. Finally, the conclusions are presented in Section \ref{Conclusions}.

\section{The system of equations}\label{SystemOfEquations}

In this section we briefly review the charged TOV equations and the physical criteria we want to use to analyze the solutions.

We consider static spherically symmetric solutions of the Einstein-Maxwell system of equations with perfect fluid as matter model. We use geometrized units, $G=c=1$, and $\epsilon_0=(4\pi)^{-1}$. The system of equations is
\begin{equation} 
    G_{\mu\nu}= 8 \pi T_{\mu\nu} , \quad \nabla _{\nu} F^{\mu \nu}= 4\pi j^{\mu}
\end{equation}
where $G_{\mu\nu}$ is the Einstein tensor, $T_{\mu\nu}$ the energy-momentum tensor, $F_{\mu\nu}$ the Faraday tensor, $j^\mu$  the electric current and $\nabla_\mu$ the covariant derivative. In this case
\begin{align}
    T_{\mu\nu} = & (\epsilon+p) u_{\mu} u_{\nu} + p g_{\mu\nu} \\
    & + \frac{1}{4\pi} \left(F_{\gamma\mu} F^{\gamma}\,_{\nu}-\frac{1}{4}F_{\gamma\lambda}F^{\gamma\lambda}g_{\mu\nu}\right), 
\end{align}
where $\epsilon$ is the energy density of the fluid, $p$ its pressure and $u^\mu$ its four-velocity field. The current $j^\mu$ is
\begin{equation}
    j^\mu = \sigma u^\mu
\end{equation}
where $\sigma$ is the charge density.

The metric in Schwarzschild coordinates takes the form
\begin{equation} \label{Metric tensor}
 ds^2 = -e^{2\Phi(r)}dt^2 + e^{2\Lambda(r)}dr^2 + r^2(d\theta^2 + \sin^2\theta \, d\phi^2).
\end{equation}
To describe the matter content in a static and spherically symmetric configuration we need the matter energy density $\epsilon(r)$, pressure $p(r)$ and charge density $\sigma(r)$. To simplify the system of equations it is convenient to define two functions, the charge inside a sphere of radius $r$, $q(r)$,
\begin{equation} \label{Maxwell-Gauss}
 q = \int_0^r 4\pi r^2 e^\Lambda \sigma\,dr,
\end{equation}
and the function $m(r)$, corresponding to the gravitational mass,
\begin{equation}
 m = \frac{r}{2}\left(1-e^{-2\Lambda}+\frac{q^2}{r^2}\right).
\end{equation}
Then the Einstein-Maxwell equations reduce to
\begin{align}
 & \frac{dm}{dr} = 4\pi r^2\left(\epsilon + \frac{\sigma q}{\sqrt{r^2-2mr+q^2}}\right), \\
 & \frac{dq}{dr} = \frac{4\pi r^3 \sigma}{\sqrt{r^2-2mr+q^2}}, \\
 & \frac{dp}{dr} = \frac{\sigma q}{r\sqrt{r^2-2mr+q^2}} - (\epsilon + p)\frac{4\pi r^3 p + m - q^2/r}{r^2 - 2 m r + q^2},
\end{align}
which are the TOV equations generalized for the presence of electric charge. The initial conditions for the ODE system are $m(r=0)=0$, $q(r=0)=0$, $p(r=0)=p_0$, where $p_0$ is arbitrary and is used as parameter in a given family of solutions. In order to close the system of equations we need constitutive relations for the fluid. As equation of state we take a relativistic polytrope, which relates the pressure to the rest-mass density, $\rho$,
\begin{equation}\label{polytropicEOS}
 p = \kappa \rho^\gamma,
\end{equation}
and then the energy density is
\begin{equation}
 \epsilon = \rho + \frac{\kappa}{\gamma-1} \rho^\gamma.
\end{equation}
For the charge density, we consider that there is a microscopically given and constant ratio, $\alpha$, between charge and rest mass,
\begin{equation}
 \sigma = \alpha\rho.
\end{equation}
As we integrate $p$ in the TOV system, we write
\begin{equation}\label{eos}
 \epsilon = \left(\frac{p}{\kappa}\right)^\frac{1}{\gamma} +\frac{p}{\gamma-1},\quad \sigma=\alpha\left(\frac{p}{\kappa}\right)^\frac{1}{\gamma}.
\end{equation}
We see that to find a solution of the system of equations, what we call an ``object'', we need to provide four constants, two corresponding to the equation of state, $\kappa$ and $\gamma$, one corresponding to the fundamental charge-mass relation of the fluid, $\alpha$, and the central pressure of the object, $p_0$. Therefore, we consider that $p_0$ parameterize a family of objects made of ``the same fluid''. Each object in the family represents a ``compact object'' in the sense that it has a finite radius, $R$, given by the condition that the pressure becomes zero, $p(R)=0$. Then, the exterior solution, $r>R$, is the RN solution with mass $M=m(R)$ and charge $Q=q(R)$. We will follow the customary criteria that there is a change of stability in a given family once the first maximum in $M$ is found while increasing $p_0$.

The speed of sound in the fluid is $c_s = \sqrt{\partial p/\partial\epsilon}$ and the causality condition is $c_s < 1$. In our case, we consider $\gamma>\frac{6}{5}$, and then $c_s < 1$ is automatically satisfied for $\frac{6}{5}<\gamma \leq 2$. For $\gamma>2$, in order for causality to be respected we need that
\begin{equation}\label{condP}
 p < \kappa^{\frac{1}{1-\gamma}}\left(\frac{\gamma-1}{\gamma(\gamma-2)}\right)^{\frac{\gamma}{\gamma-1}}.
\end{equation}
Once $\kappa$ and $\gamma$ have been fixed, then $\eqref{condP}$ gives a condition on the acceptable pressures within the object. In case we found a solution for which \eqref{condP} is violated, then we have to discard said solution as unphysical.
If instead of \eqref{polytropicEOS} we take $p=\kappa\epsilon^\gamma$ then the causality condition is $p < \kappa^\frac{1}{1-\gamma}\gamma^\frac{\gamma}{1-\gamma}$, which is more restrictive than \eqref{condP}. This ensures that we are not arbitrarily restricting the space of solutions.

In order to numerically integrate the TOV equations we need to use dimensionless variables. As we use geometrized units, we can choose a unit of length or mass and express all involved quantities in these units. For simplicity we choose a solar mass, $M_\odot = 1.9885\times 10^{30}\, kg= 1.48\,km$, as unit of mass and length. Then, the system of equations has the same form, were we have performed the replacements $r \ra M_\odot r$, $M \ra M_\odot M$, $Q \ra M_\odot Q$, $\epsilon \ra M_\odot^{-2} \epsilon$, $\sigma \ra M_\odot^{-2} \sigma$, $p \ra M_\odot^{-2} p$. Also, the equation of state and the charge-mass relation are given by \eqref{eos} with $\kappa \ra M_\odot^{2(\gamma-1)}\kappa$.

As we integrate the system of equations for different values of $\kappa$ and $\gamma$, it is necessary to take these parameters in a way that is meaningful for the intended use, namely, to study possible bounds on extremality. The existing objects that most closely motivate this search are neutron stars. Therefore, we take as a representative central density and central pressure values in the range expected for neutron stars, that is,
\begin{equation}
 \rho_{ns} = 10^{18}\,\frac{kg}{m^3},\quad p_{ns} = 10^{34}\,Pa.
\end{equation}
The dimensionless values are
\begin{equation}
 \rho_{ns}= 1.62 \times 10^{-3},\quad p_{ns} = 1.80 \times 10^{-4}.
\end{equation}
Then, having chosen a value for $\gamma$, the corresponding representative value for $\kappa$ is
\begin{equation}\label{kappaRep}
 \kappa = \frac{p_{ns}}{\rho_{ns}^\gamma}.
\end{equation}

Besides analyzing the general behaviour of charged polytropic solutions our interest is to study the possibility of black hole mimickers. Therefore we need to consider extremality criteria. As the interior solution is continued with an exterior RN spacetime, the extremality criteria corresponds with the RN black hole. Regarding the macroscopic charge, $Q$, the object is extreme if $Q=M$, sub-extreme if $Q<M$ and over-extreme if $Q>M$. Regarding how compact the object is, we compare the radius $R$ to the corresponding horizon radius, $R_+ = M+\sqrt{M^2-Q^2}$, and define the compactness parameter $C = \frac{R_+}{R}$. If $C\ra 1$, then the object would be a black hole mimicker, as it would be observationally indistinguishable from a black hole.

\section{The families of solutions}\label{FamiliesOfSolutions}

\begin{figure*}[t]
    \centering
    \begin{subfigure}[t]{0.3\textwidth}
        \centering
        \includegraphics[width=\linewidth]{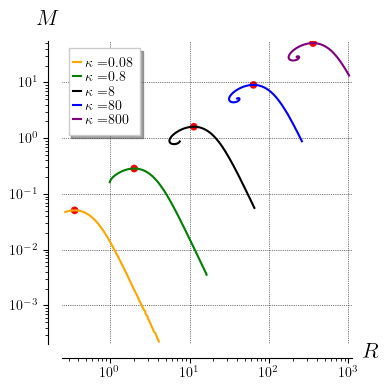} 
    \end{subfigure}
    \hfill
    \begin{subfigure}[t]{0.3\textwidth}
        \centering
        \includegraphics[width=\linewidth]{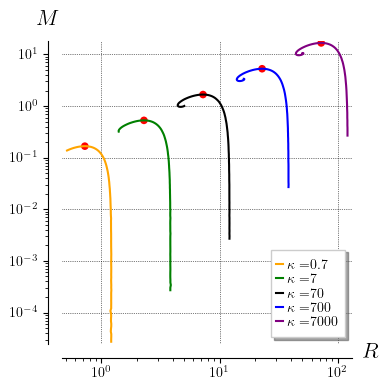} 
    \end{subfigure}
    \hfill
    \begin{subfigure}[t]{0.3\textwidth}
        \centering
        \includegraphics[width=\linewidth]{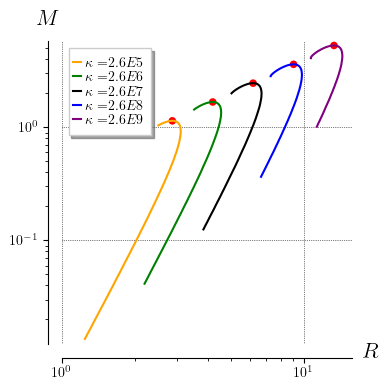} 
    \end{subfigure}

    \begin{subfigure}[t]{0.3\textwidth}
        \centering
        \includegraphics[width=\linewidth]{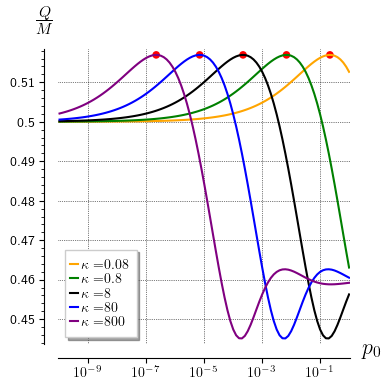} 
    \end{subfigure}
    \hfill
    \begin{subfigure}[t]{0.3\textwidth}
        \centering
        \includegraphics[width=\linewidth]{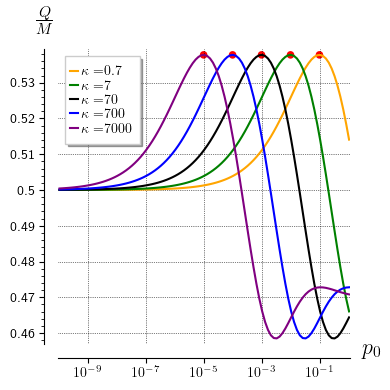} 
    \end{subfigure}
    \hfill
    \begin{subfigure}[t]{0.3\textwidth}
        \centering
        \includegraphics[width=\linewidth]{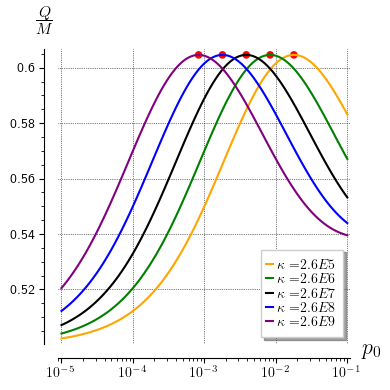} 
    \end{subfigure}

    \begin{subfigure}[t]{0.3\textwidth}
        \centering
        \includegraphics[width=\linewidth]{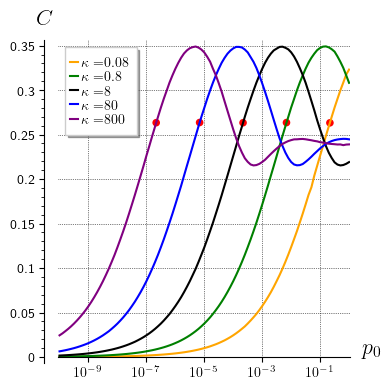} 
        \caption{$\gamma=\frac{5}{3}$} 
    \end{subfigure}
    \hfill
    \begin{subfigure}[t]{0.3\textwidth}
        \centering
        \includegraphics[width=\linewidth]{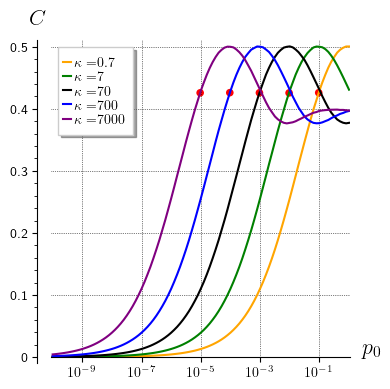} 
        \caption{$\gamma=2$} 
    \end{subfigure}
    \hfill
    \begin{subfigure}[t]{0.3\textwidth}
        \centering
        \includegraphics[width=\linewidth]{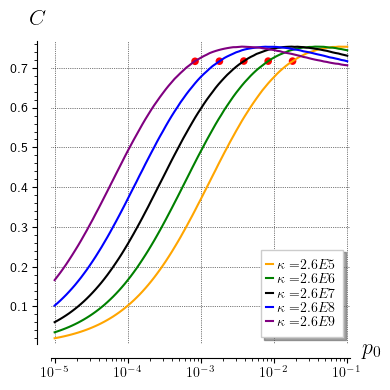} 
        \caption{$\gamma=4$} 
    \end{subfigure}

    \caption{$M$ vs. $R$, $\frac{Q}{M}$ vs. $p_0$ and $C$ vs. $p_0$ for families with varying $\kappa$. In all cases $\alpha=0.5$ and each column corresponds to a given value of $\gamma$. The red dots indicate the last stable solution.}
    \label{graficosKappa}
\end{figure*}

\begin{figure*}[t]
    \centering
    \begin{subfigure}[t]{0.3\textwidth}
        \centering
        \includegraphics[width=\linewidth]{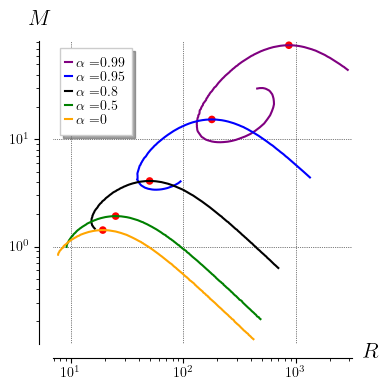} 
    \end{subfigure}
    \hfill
    \begin{subfigure}[t]{0.3\textwidth}
        \centering
        \includegraphics[width=\linewidth]{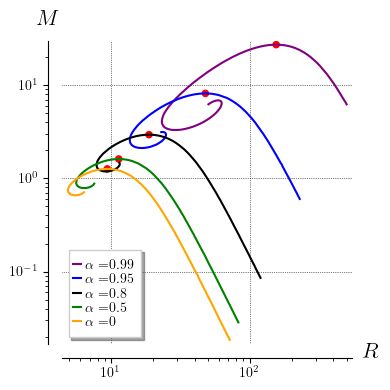} 
    \end{subfigure}
    \hfill
    \begin{subfigure}[t]{0.3\textwidth}
        \centering
        \includegraphics[width=\linewidth]{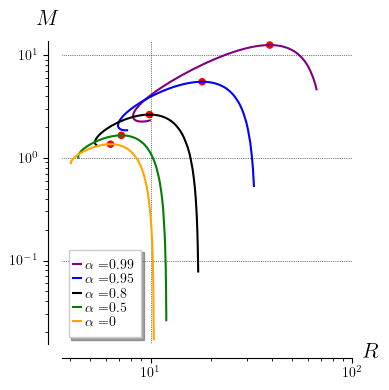} 
    \end{subfigure}

    \begin{subfigure}[t]{0.3\textwidth}
        \centering
        \includegraphics[width=\linewidth]{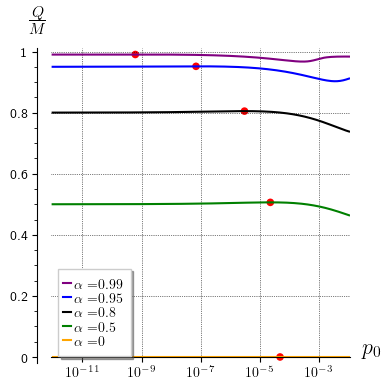} 
    \end{subfigure}
    \hfill
    \begin{subfigure}[t]{0.3\textwidth}
        \centering
        \includegraphics[width=\linewidth]{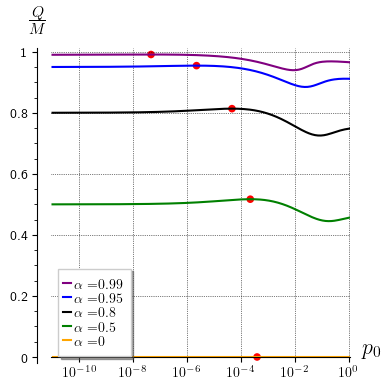} 
    \end{subfigure}
    \hfill
    \begin{subfigure}[t]{0.3\textwidth}
        \centering
        \includegraphics[width=\linewidth]{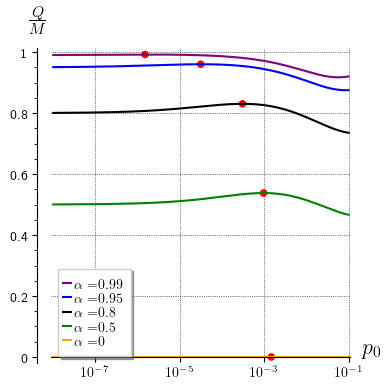} 
    \end{subfigure}

    \begin{subfigure}[t]{0.3\textwidth}
        \centering
        \includegraphics[width=\linewidth]{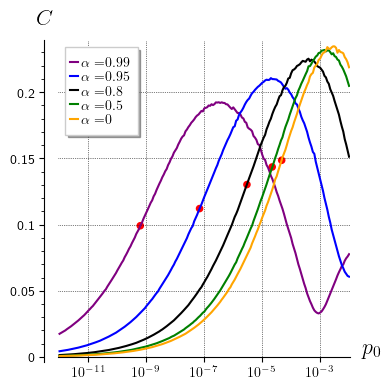} 
        \caption{$\gamma=\frac{3}{2}$} 
    \end{subfigure}
    \hfill
    \begin{subfigure}[t]{0.3\textwidth}
        \centering
        \includegraphics[width=\linewidth]{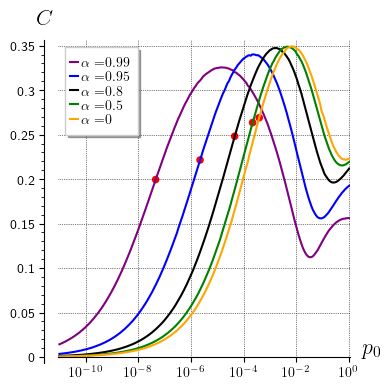} 
        \caption{$\gamma=\frac{5}{3}$} 
    \end{subfigure}
    \hfill
    \begin{subfigure}[t]{0.3\textwidth}
        \centering
        \includegraphics[width=\linewidth]{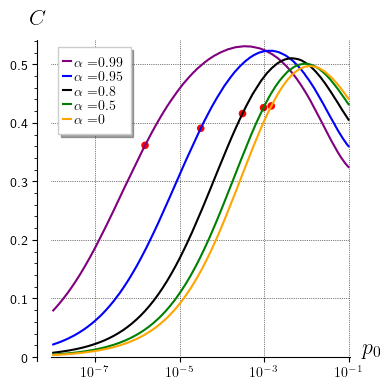} 
        \caption{$\gamma=2$} 
    \end{subfigure}

    \caption{$M$ vs. $R$, $\frac{Q}{M}$ vs. $p_0$ and $C$ vs. $p_0$ for families with varying $\alpha$. Each column corresponds to a given value of $\gamma$. The red dots indicate the last stable solution.}
    \label{graficosAlphaSoft}
\end{figure*}

\begin{figure*}[t]
    \centering
    \begin{subfigure}[t]{0.3\textwidth}
        \centering
        \includegraphics[width=\linewidth]{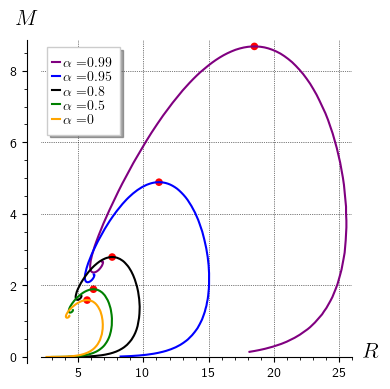} 
    \end{subfigure}
    \hfill
    \begin{subfigure}[t]{0.3\textwidth}
        \centering
        \includegraphics[width=\linewidth]{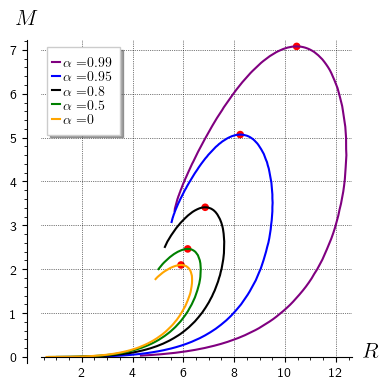} 
    \end{subfigure}
    \hfill
    \begin{subfigure}[t]{0.3\textwidth}
        \centering
        \includegraphics[width=\linewidth]{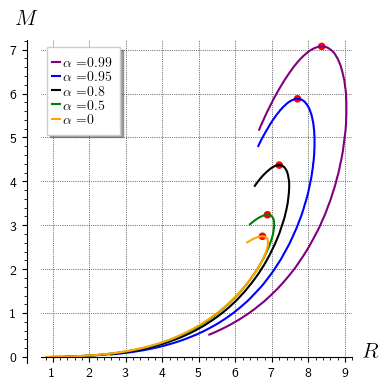} 
    \end{subfigure}

    \begin{subfigure}[t]{0.3\textwidth}
        \centering
        \includegraphics[width=\linewidth]{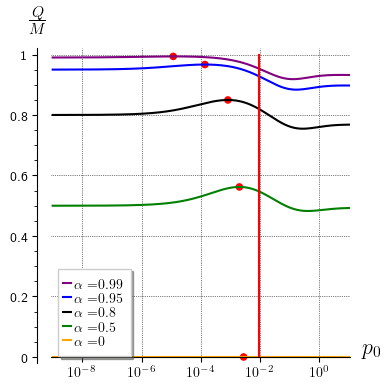} 
    \end{subfigure}
    \hfill
    \begin{subfigure}[t]{0.3\textwidth}
        \centering
        \includegraphics[width=\linewidth]{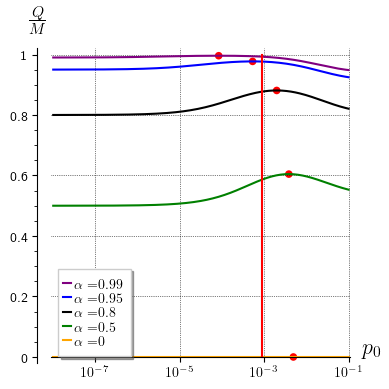} 
    \end{subfigure}
    \hfill
    \begin{subfigure}[t]{0.3\textwidth}
        \centering
        \includegraphics[width=\linewidth]{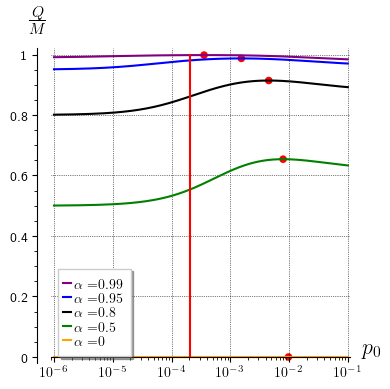} 
    \end{subfigure}

    \begin{subfigure}[t]{0.3\textwidth}
        \centering
        \includegraphics[width=\linewidth]{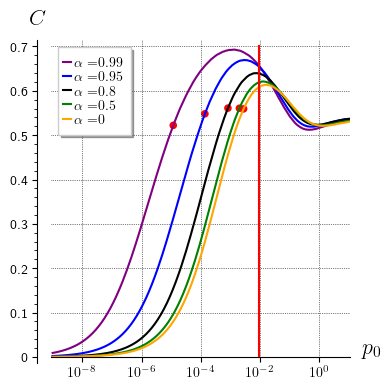} 
        \caption{$\gamma=\frac{5}{2}$} 
    \end{subfigure}
    \hfill
    \begin{subfigure}[t]{0.3\textwidth}
        \centering
        \includegraphics[width=\linewidth]{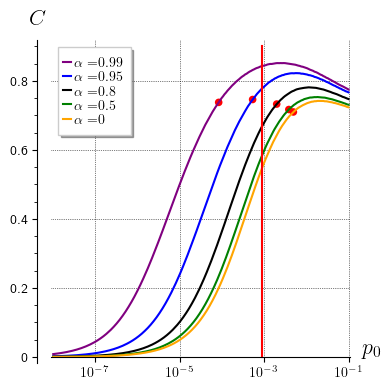} 
        \caption{$\gamma=4$} 
    \end{subfigure}
    \hfill
    \begin{subfigure}[t]{0.3\textwidth}
        \centering
        \includegraphics[width=\linewidth]{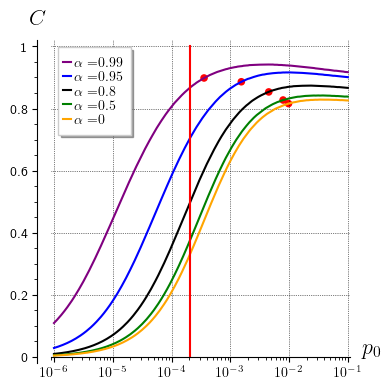} 
        \caption{$\gamma=9$} 
    \end{subfigure}

    \caption{$M$ vs. $R$, $\frac{Q}{M}$ vs. $p_0$ and $C$ vs. $p_0$ for families with varying $\alpha$. Each column corresponds to a given value of $\gamma$. The red dots indicate the last stable solution. The red lines indicate the last causal solution.}
    \label{graficosAlphaStiff}
\end{figure*}

We integrate the TOV system of equations for different values of the parameters. The profiles for the involved functions, $p$, $m$ and $q$, are all standard and well behaved, and therefore we concentrate on discussing the families of solutions. Each family corresponds to a given set of values for $\alpha$, $\kappa$ and $\gamma$, and is parameterized by $p_0$. The quantities that we want to discuss are the total mass, $M$, radius, $R$, charge to mass, $\frac{Q}{M}$, and compactness, $C$. We separate the stable solutions from the unstable ones by finding the first maximum of $M$ as a function of $p_0$.

We start considering how the families of solutions change when we vary $\kappa$. In Figure \ref{graficosKappa} we present the graphs for $M$ as a function of $R$, $\frac{Q}{M}$ as a function of $p_0$ and $C$ also as a function of $p_0$. Each column corresponds to a given value of $\gamma$, with values $\frac{5}{3}$, $2$ and $4$. In each graph there are plotted five families of solutions, each corresponding to a value of $\kappa$. The last stable solution is represented with a red dot. All graphs have $\alpha=0.5$. From the first line of plots we see that by increasing $\kappa$ the mass and radius of the solution increase, and therefore also the maximum mass increases. Modulo this displacement, all families of solutions behave identically for a given value of $\gamma$. Remarkably, for the $\frac{Q}{M}$ and $C$ relationships, the graphs seem to be simple displacements regarding $p_0$, actually, in $\log p_0$. Therefore, within numerical accuracy, the limiting values for $p_0 \ra 0$ and the maximum values do not depend on $\kappa$. We see that as expected $\lim_{p_0\ra 0} \frac{Q}{M} = \alpha$ and that $\lim_{p_0\ra 0} C = 0$. Also, the maximum value of $\frac{Q}{M}$ is above $\alpha$ but far from $1$, and the maximum value of $C$ increases with $\gamma$ but again is far from $1$. Given that varying $\kappa$ does not give new phenomenology, and that we want to concentrate on the measures of extremality, from now on, once $\gamma$ is fixed, the corresponding $\kappa$ is given by \eqref{kappaRep}, and then we have one less parameter to consider.

We now turn our attention to how the families behave with respect to $\alpha$ and $\gamma$. In Figures \ref{graficosAlphaSoft} and \ref{graficosAlphaStiff} we plot the same quantities as in Figure \ref{graficosKappa}, each column with a given value of $\gamma$, having for Figure \ref{graficosAlphaSoft} the values $\frac{3}{2}$,
$\frac{5}{3}$ and $2$, and for Figure \ref{graficosAlphaStiff} the values $\frac{5}{2}$, $4$ and $9$. In each graph, each family corresponds to a given value of $\alpha$, again being indicated by a red dot the last stable solution. For $\gamma>2$, a vertical red line indicates the separation between solutions that respect causality from those that violate it, being the acceptable solutions to the left of said line. From the $M-R$ diagrams we see that for a given $\gamma$, the solutions are more extended and massive the higher $\alpha$ is. Please note that in the first line of plots in Figure \ref{graficosAlphaSoft} the scale is $\log-\log$ while in Figure \ref{graficosAlphaStiff} it is linear. This means that for $\gamma<2$ a change in $\alpha$ produces a big change in $M$ and $R$, while for $\gamma>2$ the effect is much less pronounced. This also corresponds to how soft or stiff the equation of state is, and we can see that this does not depend on $\alpha$. We see that $\gamma=2$ is the transition value from a gas like behaviour for $\gamma<2$ to a liquid like behaviour for $\gamma>2$. For $\gamma<2$ the limit $p_0\ra 0$ has $R\ra\infty$ while for $\gamma>2$ the limit $p_0\ra 0$ has $R\ra 0$. In other words, for the gas like behaviour there is no maximum radius for a family of objects, while for the liquid like behaviour there is no minimum radius. The transition value is $\gamma=2$, where for a given family there is both a maximum and minimum radius.

Turning our attention to the ratio $\frac{Q}{M}$, the behaviour is as expected. That is, $\lim_{p_0\ra 0}\frac{Q}{M}=\alpha$, and it increases with increasing $p_0$, although we always have $\frac{Q}{M}<1$. We have tried to construct solutions with $\alpha>1$ but have been unsuccessful, even for $\alpha$ slightly above $1$. For a given family, the maximum of $\frac{Q}{M}$ is obtained for the last stable solution, but if we increase $\gamma$ the causality condition starts to limit the range of $p_0$ and said maximum can not be attained. It is interesting to note that the maximum value of $p_0$ allowed for a given family decreases with increasing $\alpha$, we explore this in the next section.

Finally, we consider $C$. As expected, for given values of $\gamma$ and $\alpha$, $C$ is an increasing function of $p_0$ and the allowed maximum is attained for the last stable object in the family. If $\gamma$ is big enough, the last acceptable object is limited by causality, and the maximum value of $C$ is obtained for the last causal object. If we now consider $\gamma$ as fixed and check the maximum values of $C$ as functions of $\alpha$ we see that the behaviour depends on the value of $\gamma$. For low $\gamma$ the maximum of $C$ is obtained for $\alpha=0$. That is, if the equation of state is soft enough, the most compact object possible corresponds to the neutral case. For $\gamma$ big enough, the behaviour reverses, and the most compact object corresponds to the biggest $\alpha$. Also, as $\gamma$ increases the maximum of $C$ approaches $C=1$, although the range of $p_0$ gets limited more and more. In the next section we explore this limit.

\section{The black hole mimicker limit}\label{MimickerLimit}

\begin{figure*}[t]
    \centering
    \begin{subfigure}[t]{0.3\textwidth}
        \centering
        \includegraphics[width=\linewidth]{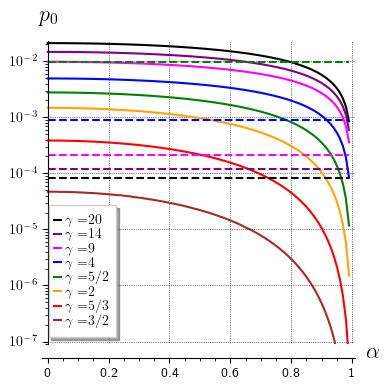} 
    \end{subfigure}
    \hfill
    \begin{subfigure}[t]{0.3\textwidth}
        \centering
        \includegraphics[width=\linewidth]{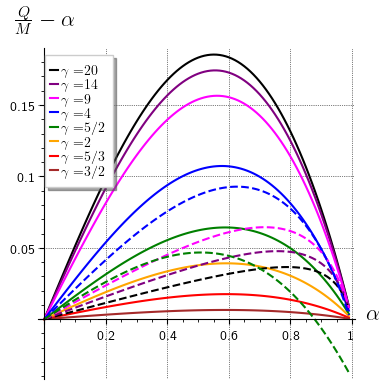} 
    \end{subfigure}
    \hfill
    \begin{subfigure}[t]{0.3\textwidth}
        \centering
        \includegraphics[width=\linewidth]{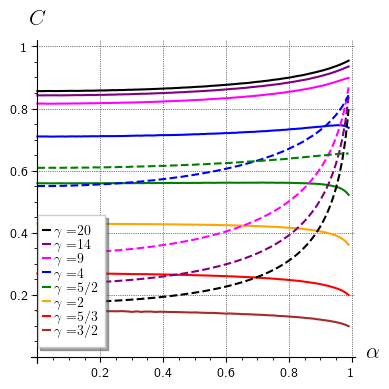} 
    \end{subfigure}

    \begin{subfigure}[t]{0.3\textwidth}
        \centering
        \includegraphics[width=\linewidth]{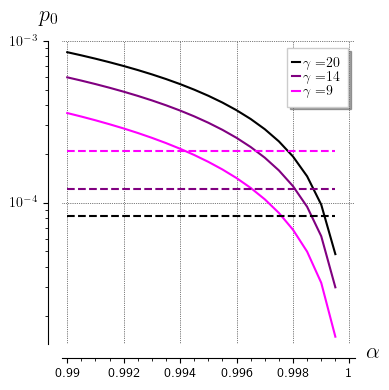} 
    \end{subfigure}
    \hfill
    \begin{subfigure}[t]{0.3\textwidth}
        \centering
        \includegraphics[width=\linewidth]{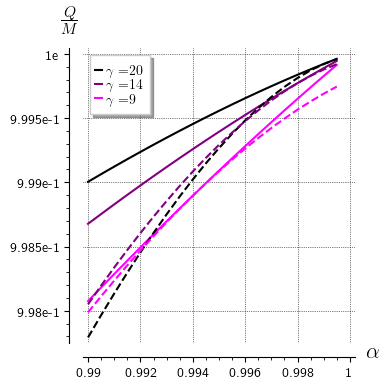} 
    \end{subfigure}
    \hfill
    \begin{subfigure}[t]{0.3\textwidth}
        \centering
        \includegraphics[width=\linewidth]{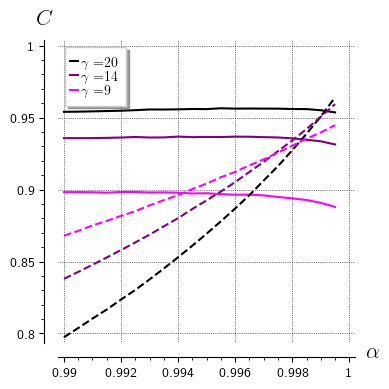} 
    \end{subfigure}

    \caption{Maximum allowed values of $p_0$, $\frac{Q}{M}$ and $C$ as functions of $\alpha$ for several values of $\gamma$. Solid lines indicate bounds due to stability and dashed lines indicate bounds due to causality.}
    \label{graficosCrit}
\end{figure*}

In this section we present the analysis of the possibility of obtaining black hole mimickers made of polytropic charged matter. For this we analyze the allowed values of $p_0$ and the extremality measures, $\frac{Q}{M}$ and $C$. In Figure \ref{graficosCrit} we present the maximum allowed values of $p_0$, $\frac{Q}{M}$ and $C$ as functions of $\alpha$ and for several values of $\gamma$. The first line of plots corresponds to $0\leq\alpha\leq0.99$ while the second line zooms into the range $0.99\leq\alpha\leq0.999$. In the graphs we have indicated with a solid line the bounds corresponding to the last stable object in the family, while the dashed line correspond to the last causal object in the same family. Therefore, for $\gamma\leq2$ there is no dashed line. We see that, for fixed $\alpha$ the allowed range for $p_0$ regarding stability is an increasing function of $\gamma$. Conversely, regarding causality, said range is a decreasing function. Also, regarding causality, we have that $\lim_{\alpha\ra 1}p_0 = 0$. This is a very strong restriction, as it shows that if we want to approach the limit $\alpha\ra 1$ we at the same time have to reduce the pressure and approach $p_0=0$. In fact, in the case $\alpha=1$ we are left with $p_0=0$, and therefore no matter is present.

Regarding $\frac{Q}{M}$, in the second column of Figure \ref{graficosCrit} we plot first $\frac{Q}{M}-\alpha$, as it is clearer the different behaviour for different values of $\gamma$. Here it is important to check first if the bound is imposed due to stability or due to causality. For example, for $\gamma=\frac{5}{2}$ the bound is only due to stability, while for $\gamma=9$ the bound is first due to causality and then, close to $\alpha =1$ it is due to stability. The main conclusion here is that $\frac{Q}{M}$ can be above $\alpha$ but that always $\frac{Q}{M}<1$ for $\alpha<1$ and that for all $\gamma$ we have $\lim_{\alpha\ra 1}\frac{Q}{M}=1$.

Lastly, considering $C$, for a given $\alpha$, regarding stability $C$ is an increasing function of $\gamma$, while regarding causality one needs to be careful, because it depends on the particular value of $\alpha$, although generally it is a decreasing function of $\gamma$. More importantly, $C$ is bounded above away from $C=1$ for any given $\gamma$. We see that if we want to approach $C=1$ we need to take two limits at the same time, $\alpha\ra 1$ and $\gamma\ra\infty$. This is our main conclusion, that to obtain a black hole mimicker we need to take said double limit, and this in fact imposes also the limit $p_0\ra 0$. Therefore, if we just take those limits we are left with an empty flat spacetime. The only workaround is to notice that the limit can be taken in such a way that at the same time $\kappa\ra 0$, and if done carefully, although $p\ra 0$ we have $\rho\neq 0$. This limit is in fact the known ECD spacetimes, for which $p=0$ and $\alpha=1$, and which correspond to an ERN exterior.

\section{Conclusions}\label{Conclusions}

We have numerically constructed and analyzed families of electrically charged polytropic spheres. Regarding the  constitutive relation for charge, we take the charge density to be proportional to rest mass density, as we consider that the fluid is composed of particles with a fixed charge to mass ratio. With respect to the equation of state, we considered relativistic polytropes, where the polytropic relation is with respect to rest mass density. To consider the obtained objects as physically feasible, we impose the usual stability condition and also the causality condition of the speed of sound being below the speed of light. Our main conclusion is that quasiblack holes can not be formed of charged polytropic matter. In fact, the limit for obtaining a black hole mimicker is subtle and ends up being ECD. This also shows that starting with polytropic matter we can not construct non-extremal black hole mimickers. It is still possible to construct charged solutions with $\frac{Q}{M}\approx 1$, but this solutions are quite extended and faint, and the closer we get to extremality the closer we are to ECD. These results lend support to the conjecture of the impossibility of forming extremal black holes through the collapse of regular charged matter.

\section{Acknowledgments}

Part of the calculations including the numerical integrations were performed using SageMath \cite{sagemath}. For the General Relativity calculations the package SageManifolds \cite{sagemanifolds} was used.



\bibliography{biblio}
\bibliographystyle{abbrvnat}

\end{document}